\pdfoutput=1
\documentclass[onefignum,onetabnum]{siamart171218}

\usepackage{amsfonts}
\usepackage{amsmath}
\usepackage{graphicx}
\usepackage{epstopdf}
\usepackage{booktabs}
\usepackage{subcaption}
\usepackage{float}
\usepackage{enumitem}


\ifpdf
  \DeclareGraphicsExtensions{.eps,.pdf,.png,.jpg}
\else
  \DeclareGraphicsExtensions{.eps}
\fi

\newcommand{\mat}[1]{\mathbf{#1}}
\newcommand{\vct}[1]{\boldsymbol{#1}}

\headers{Quantum HHL Algorithm for 1D Linear Advection Equation}{B. Yalcin and L. T. Zhang}

\title{Quantum HHL Algorithm for Solving 1D Linear Advection Equation Using Explicit and Implicit Finite Difference Schemes\thanks{This research received no external funding.}}

\author{%
  Batuhan Yalcin\thanks{PhD Student, Department of Mechanical, Aerospace and Nuclear Engineering (MANE), Rensselaer Polytechnic Institute (RPI), Troy, NY 12180 (yalcib@rpi.edu).}
  \and
  Lucy T. Zhang\thanks{Professor, Department of Mechanical, Aerospace and Nuclear Engineering (MANE), Rensselaer Polytechnic Institute (RPI), JEC5010 (mail stop JEC2049-MANE), 110 8th St., Troy, NY 12180 (zhangl12@rpi.edu). \url{https://faculty.rpi.edu/lucy-zhang}.}}

\begin{document}

\maketitle

\begin{abstract}
This study develops and implements a Harrow-Hassidim-Lloyd (HHL) quantum algorithm for solving a one-dimensional linear advection equation with periodic boundary conditions. The governing equation is first discretized using a classical finite difference (FD) scheme to obtain a linear system suitable for quantum solution. We present a newly designed, circuit-level implementation of HHL assembled from Qiskit circuit components and exploit the circulant structure of the periodic discretization matrices through Fourier-based matrix evolution. 
The HHL algorithm is benchmarked with three explicit schemes (Explicit Upwind Differencing, Lax-Friedrichs, and Lax-Wendroff) and one implicit scheme (Crank-Nicolson). The implicit scheme requires a nontrivial linear solve arising from the discretized PDE on progressively finer grids, whereas the explicit schemes provide different input fields for testing state preparation and reconstruction with identity coefficient matrices. HHL solution states are validated against direct classical solves in noiseless simulation. For hardware execution, the periodic matrix structure permits a Fourier implementation that applies reciprocal eigenvalues without phase estimation. Results on the IBM Rensselaer Nighthawk processor are compared at a shared final time. Noiseless reconstructed fields approach the transported profile on resolved grids, while substantial hardware errors remain for both explicit and implicit schemes. Crank-Nicolson remains exceptionally well-conditioned and utilizes similar circuit resources to the explicit controls, proving that hardware degradation in our tests was not driven by poor matrix conditioning. Ultimately, this study shows new means for HHL algorithms to study quantum CFD problems involving advection equations with implicit time integrations, while demonstrating how state preparation and measurement affect the practical performance of quantum solvers on current devices.
\end{abstract}



\section{Introduction}
Partial Differential Equations (PDEs) are the mathematical backbone of engineering physics, governing phenomena ranging from fluid dynamics (Navier-Stokes equations) to electromagnetism (Maxwell's equations) and heat transfer. In high-fidelity Computational Fluid Dynamics (CFD), the numerical solution of these equations via classical methods, such as Finite Difference (FD), Finite Element (FE), and Finite Volume (FV), often becomes computationally prohibitive as grid resolution increases \cite{ref:classical_pde, ref:cfd_review}.

Quantum computing offers a paradigm shift for high-performance computing. Specifically, the Harrow--Hassidim--Lloyd (HHL) algorithm \cite{ref:hhl} promises an exponential speedup for solving systems of linear equations of the form $\mat{A}\vct{x} = \vct{b}$. Since implicit numerical schemes for PDEs reduce to solving massive linear systems at every time step, HHL is a prime candidate for accelerating CFD simulations \cite{ref:montanaro_pde, ref:gaitan_cfd}. While classical algorithms typically scale as $O(N)$ or $O(N \log N)$ where $N$ is the matrix dimension, HHL scales theoretically as $O(\log N \cdot \kappa^2 \cdot \epsilon^{-1})$, where $\kappa$ is the condition number of the matrix and $\epsilon$ is the desired precision. Newer approaches, such as those by Childs et al. \cite{ref:childs2017}, have further improved the dependence on precision, though hardware implementation remains challenging. Realizing this advantage on current Noisy Intermediate-Scale Quantum (NISQ) hardware is non-trivial \cite{ref:preskill_nisq}: limitations in qubit coherence times, gate fidelity, and connectivity often introduce additional costs and errors.

In this study, we apply quantum linear-solver constructions to the 1D linear advection equation, a hyperbolic PDE representing wave propagation. We compare classical FD schemes on a fine grid ($N_{full}=128$) with quantum reconstructions on coarse grids ($N_c$). For the explicit schemes, the identity coefficient matrix makes inversion trivial; their different numerical dissipation and dispersion provide three distinct right-hand sides for testing state preparation and reconstruction. Crank--Nicolson supplies the nontrivial implicit solve. The algorithmic and implementation contributions are:\\
1) a common formulation of the explicit updates and the Hermitian reformulation of the implicit CN system, \\
2) circuit-level HHL implementation with phase-parameter selection and Fourier-based evolution, validated against direct classical solves, and \\
3) a clock-free Fourier implementation with signed-field and norm reconstruction, evaluated using the \texttt{ibm\_rensselaer} measurements and recorded circuit resources. 
Although HHL and Fourier diagonalization are established methods \cite{ref:hhl,ref:zhou}, our contribution is their implementation and evaluation for the explicit and implicit periodic-advection systems considered here. We validate our circuit-level HHL implementation in noiseless simulation and evaluate the separate clock-free Fourier construction on quantum hardware. The comparison examines reconstructed-field accuracy and recorded circuit resources
without claiming an algorithmic advantage for the explicit schemes.

The rest of the paper is organized as follows. In Section \ref{sec:methodology}, we describe the governing equation and finite difference discretization schemes. Linear systems of equations are established for three explicit (Explicit Upwind Differencing, Lax-Friedrichs, Lax-Wendroff) and one implicit (Crank-Nicolson) schemes. In Section \ref{sec:HHL}, our implemented HHL algorithm for the four discretization schemes is presented. The handling of non-Hermitian systems for Crank-Nicolson and the circulant structure for periodic boundary conditions are described in detail. The noiseless HHL validation results and the experimental results of our clock-free Fourier solver executed on real IBM quantum
hardware (\texttt{ibm\_rensselaer}) are presented in Section~\ref{sec:results}, followed by a discussion in Section~\ref{sec:discussion}. Finally, we conclude with our findings in Section \ref{sec:conclusions}.

\section{Methodology}\label{sec:methodology}

\subsection{Governing Equation and Discretization}
We consider 1D linear advection equation, which describes the transport of a scalar quantity $\phi$ with constant velocity $u$:
\begin{equation}
    \frac{\partial \phi}{\partial t} + u \frac{\partial \phi}{\partial x} = 0
    \label{eq:pde}
\end{equation}
The domain is defined as $x \in [0, L]$ with periodic boundary conditions $\phi(0, t) = \phi(L, t)$. The simulation parameters are length $L=2.0$, velocity $u=0.2$, and Courant number $\alpha = \frac{u \Delta t}{\Delta x} = 0.8$. The simulation runs until time $T_s=0.5$.

The initial condition, representing a distinct wave packet, is given by:
\begin{equation}
    \phi(x, 0) =
    \begin{cases}
        \sin(4 \pi (x - 0.5)) & \text{if } 0.5 \le x \le 1.5 \\
        0 & \text{otherwise}
    \end{cases}
    \label{eq:initial_condition}
\end{equation}

The exact solution is the periodic translation $\phi(x,t)=\phi((x-ut)\bmod L,0)$. There is no physical diffusion in this equation. Therefore, the damping in the FD solutions is purely numerical. In the quantum benchmark, preceding time steps are advanced classically and the circuit solves the final-step linear system using the resulting right-hand side. The experiment therefore tests one discrete solve and its readout, rather than an entirely quantum time-marching calculation.

\subsection{Classical Finite Difference Schemes}
To discretize Eq.\ (\ref{eq:pde}), we use a fine grid with $N_{full}=128$ points at $x_i=(i+1/2)L/128$, $i=0,\ldots,127$. Each coarse grid retains every $128/N_c$-th fine-grid point, with spacing $\Delta x=L/N_c$ and the same offset $L/256$. Time levels are $t^n=n\Delta t$, and spatial subscripts in the following stencils are interpreted periodically. We benchmark four distinct schemes---Explicit Upwind Differencing, Lax--Friedrichs, Lax--Wendroff, and Crank--Nicolson---as defined below.

\subsubsection{Explicit Upwind Differencing (EUD)}
This first-order scheme uses backward differencing in space (for $u>0$). It is stable for $\alpha \le 1$ but introduces significant numerical dissipation (smearing):
\begin{equation}
    \phi_i^{n+1} = \phi_i^n - \alpha (\phi_i^n - \phi_{i-1}^n)
    \label{eq:eud}
\end{equation}
To formulate this for a quantum linear solver $\mat{A}_{EUD} \vct{\phi}^{n+1} = \vct{b}_{EUD}^n$, the matrix $\mat{A}_{EUD}$ and right-hand side vector $\vct{b}_{EUD}^n$ under periodic boundary conditions are:
\begin{equation}
    \mat{A}_{EUD} = \begin{bmatrix}
    1 & 0 & \cdots & 0 \\
    0 & 1 & \cdots & 0 \\
    \vdots & \vdots & \ddots & \vdots \\
    0 & 0 & \cdots & 1
    \end{bmatrix}, \quad
    \vct{b}_{EUD}^n = \begin{bmatrix}
    \phi_1^n - \alpha(\phi_1^n - \phi_N^n) \\
    \phi_2^n - \alpha(\phi_2^n - \phi_1^n) \\
    \vdots \\
    \phi_N^n - \alpha(\phi_N^n - \phi_{N-1}^n)
    \end{bmatrix}
    \label{eq:eud_matrix}
\end{equation}

\subsubsection{Lax--Friedrichs (LF)}
The LF scheme averages the spatial term to improve stability, though it remains first-order accurate:
\begin{equation}
    \phi_i^{n+1} = \frac{1}{2}(\phi_{i+1}^n + \phi_{i-1}^n) - \frac{\alpha}{2}(\phi_{i+1}^n - \phi_{i-1}^n)
    \label{eq:lf}
\end{equation}
The corresponding linear system $\mat{A}_{LF} \vct{\phi}^{n+1} = \vct{b}_{LF}^n$ is defined by:
\begin{equation}
    \mat{A}_{LF} = \begin{bmatrix}
    1 & 0 & \cdots & 0 \\
    0 & 1 & \cdots & 0 \\
    \vdots & \vdots & \ddots & \vdots \\
    0 & 0 & \cdots & 1
    \end{bmatrix}, \quad
    \vct{b}_{LF}^n = \begin{bmatrix}
    \frac{1}{2}(\phi_2^n + \phi_N^n) - \frac{\alpha}{2}(\phi_2^n - \phi_N^n) \\
    \frac{1}{2}(\phi_3^n + \phi_1^n) - \frac{\alpha}{2}(\phi_3^n - \phi_1^n) \\
    \vdots \\
    \frac{1}{2}(\phi_1^n + \phi_{N-1}^n) - \frac{\alpha}{2}(\phi_1^n - \phi_{N-1}^n)
    \end{bmatrix}
    \label{eq:lf_matrix}
\end{equation}

\subsubsection{Lax--Wendroff (LW)}
To reduce diffusion, the LW scheme expands the Taylor series to second order, resulting in a more accurate but dispersive scheme:
\begin{equation}
    \phi_i^{n+1} = \phi_i^n - \frac{\alpha}{2}(\phi_{i+1}^n - \phi_{i-1}^n) + \frac{\alpha^2}{2}(\phi_{i+1}^n - 2\phi_i^n + \phi_{i-1}^n)
    \label{eq:lw}
\end{equation}
In matrix form $\mat{A}_{LW} \vct{\phi}^{n+1} = \vct{b}_{LW}^n$, the system is:
\begin{equation}
    \mat{A}_{LW} = \begin{bmatrix}
    1 & 0 & \cdots & 0 \\
    0 & 1 & \cdots & 0 \\
    \vdots & \vdots & \ddots & \vdots \\
    0 & 0 & \cdots & 1
    \end{bmatrix}, \quad
    \vct{b}_{LW}^n = \begin{bmatrix}
    \phi_1^n - \frac{\alpha}{2}(\phi_2^n - \phi_N^n) + \frac{\alpha^2}{2}(\phi_2^n - 2\phi_1^n + \phi_N^n) \\
    \phi_2^n - \frac{\alpha}{2}(\phi_3^n - \phi_1^n) + \frac{\alpha^2}{2}(\phi_3^n - 2\phi_2^n + \phi_1^n) \\
    \vdots \\
    \phi_N^n - \frac{\alpha}{2}(\phi_1^n - \phi_{N-1}^n) + \frac{\alpha^2}{2}(\phi_1^n - 2\phi_N^n + \phi_{N-1}^n)
    \end{bmatrix}
    \label{eq:lw_matrix}
\end{equation}

\subsubsection{Crank--Nicolson (CN)}
This is an implicit, second-order scheme that is neutrally stable and non-dissipative for linear advection. It averages the spatial derivatives at time steps $n$ and $n+1$:
\begin{equation}
    \phi_i^{n+1} + \frac{\alpha}{4}(\phi_{i+1}^{n+1} - \phi_{i-1}^{n+1}) = \phi_i^n - \frac{\alpha}{4}(\phi_{i+1}^n - \phi_{i-1}^n)
    \label{eq:cn}
\end{equation}
Eq. (\ref{eq:cn}) creates a cyclic tridiagonal system of linear equations $\mat{A}_{CN} \vct{\phi}^{n+1} = \vct{b}_{CN}^n$ that must be solved at every step. Incorporating periodic boundary conditions, the matrix $\mat{A}_{CN}$ and the right-hand side vector $\vct{b}_{CN}^n$ are explicitly structured as:
\begin{equation}
    \mat{A}_{CN} = \begin{bmatrix}
    1 & \frac{\alpha}{4} & 0 & \cdots & -\frac{\alpha}{4} \\
    -\frac{\alpha}{4} & 1 & \frac{\alpha}{4} & \cdots & 0 \\
    0 & -\frac{\alpha}{4} & 1 & \ddots & \vdots \\
    \vdots & \vdots & \ddots & \ddots & \frac{\alpha}{4} \\
    \frac{\alpha}{4} & 0 & \cdots & -\frac{\alpha}{4} & 1
    \end{bmatrix}, \quad
    \vct{b}_{CN}^n = \begin{bmatrix}
    \phi_1^n - \frac{\alpha}{4}(\phi_2^n - \phi_N^n) \\
    \phi_2^n - \frac{\alpha}{4}(\phi_3^n - \phi_1^n) \\
    \vdots \\
    \phi_N^n - \frac{\alpha}{4}(\phi_1^n - \phi_{N-1}^n)
    \end{bmatrix}
    \label{eq:cn_matrix}
\end{equation}
The periodic CN matrix is only mildly conditioned for this experiment. Its squared singular values are $1+(\alpha^2/4)\sin^2(2\pi k/N_c)$. For the powers-of-two grids considered here,
\begin{equation}
\kappa_2(\mat A_{CN})=\sqrt{1+\alpha^2/4},\qquad
\kappa_2(\mat A_{CN}^{T}\mat A_{CN})=1+\alpha^2/4.
\label{eq:conditioning}
\end{equation}
At fixed $\alpha=0.8$, the condition numbers $\kappa_2(\mat A_{CN})$ and $\kappa_2(\mat A_{CN}^{T}\mat A_{CN})$
are approximately $1.077$ and $1.16$, respectively, independent of refinement over the powers-of-two grids considered. Consequently, increasing ill-conditioning with grid refinement cannot explain the hardware errors in this fixed-Courant test.

\section{Quantum HHL Algorithm Framework} \label{sec:HHL}
\subsection{Brief Review of the Quantum HHL Algorithm}
We construct the HHL circuits in this work rather than call a packaged Qiskit HHL solver. Our code assembles state preparation, controlled matrix evolution, phase estimation, reciprocal rotation, and uncomputation, and selects the phase-estimation parameters. It uses Qiskit library components for state preparation, QFT, and reciprocal rotation; these components are not new algorithms introduced here. Qiskit supplies the circuit-building, simulation, transpilation, and hardware-execution tools for our implementation.

The core task is to solve $\mat{A} \vct{x} = \vct{b}$ for the vector $\vct{x}$. In the context of our explicit schemes, $\mat{A}$ is exactly the identity matrix $\mat{I}$. For the implicit CN scheme, $\mat{A}$ is the cyclic tridiagonal matrix defined in Eq. (\ref{eq:cn_matrix}).

The standard HHL algorithm \cite{ref:hhl} provides a method to solve linear systems by finding a quantum state $|x\rangle$ proportional to the solution vector $\vct{x}$. The implementation considered here assumes $\mat{A}$ is Hermitian positive definite and that the right-hand side can be efficiently encoded into a quantum state $|b\rangle = \sum_{j} \beta_j |u_j\rangle$, where $|u_j\rangle$ are the eigenvectors of $\mat{A}$ with corresponding eigenvalues $\lambda_j$. The evolution parameter $\tau$ is distinct from the PDE time step. In the ideal exact-phase description, the algorithm proceeds in three main phases. 1) \textbf{Quantum Phase Estimation (QPE):} The unitary operator $\mat{U} = e^{i \mat{A} \tau}$ is applied conditionally by a register of clock (ancilla) qubits to the input state $|b\rangle$. This extracts the eigenvalues $\lambda_j$ of the matrix $\mat{A}$ and stores them in the ancilla register in binary representation, transforming the state into a superposition of eigenvectors entangled with their corresponding eigenvalues: $\sum_j \beta_j |\lambda_j\rangle |u_j\rangle$. 2) \textbf{Controlled Rotation:} An additional auxiliary qubit, initially in the state $|0\rangle$, is rotated by an angle $\theta = 2 \arcsin(C/\lambda_j)$, conditioned on the eigenvalue register. This step effectively performs the matrix inversion operation $\lambda_j^{-1}$, mapping the state to $\sum_j \beta_j |\lambda_j\rangle |u_j\rangle \left( \sqrt{1-\frac{C^2}{\lambda_j^2}}|0\rangle + \frac{C}{\lambda_j}|1\rangle \right)$, where $C$ is a normalization constant chosen to be smaller than the minimum eigenvalue in magnitude ($C \le \min |\lambda_j|$). 3) \textbf{Inverse QPE:} The phase estimation circuit is applied in reverse (adjoint) to uncompute the eigenvalue register, disentangling the clock qubits and returning them to $|0\rangle$. Upon measuring the auxiliary qubit in the $|1\rangle$ state, the remaining system collapses into the state proportional to $\mat{A}^{-1}|b\rangle$, yielding the normalized solution to the linear system. 

With a finite clock register, phase estimation is approximate and residual clock entanglement can affect the solution. We therefore validate the complete circuit in statevector simulation, retaining the ancilla-$1$, clock-$0$ branch and comparing its normalized state with a direct classical solve. For the recorded CN test at $N_c=16$, increasing the clock from two to six qubits, with the evolution parameter selected for each clock size, raises the fidelity from approximately $0.4499$ to $0.99999$. State fidelity is distinct from the field-error metric used below. Circuit designs and small experimental demonstrations of HHL provide relevant precedents \cite{ref:cao_quantum_sim,ref:cai,ref:pan}.

Each phase has a distinct role in this HHL construction. Quantum Phase Estimation is used because the inversion $\lambda_j^{-1}$ cannot be applied directly to the unknown eigenbasis of $\mat{A}$; the eigenvalues must first be written into an addressable ancilla register where each eigencomponent can be operated on independently. The controlled rotation then carries out the actual inversion by weighting each eigencomponent by $1/\lambda_j$, which is the step that solves the linear system. Finally, the inverse QPE is necessary to disentangle the clock register from the solution. Without it, the measured state would remain correlated with the eigenvalue information rather than collapsing onto the normalized solution $\mat{A}^{-1}\vct{b}$ alone.

\subsection{Handling Non-Hermitian Systems}
The standard HHL algorithm requires $\mat{A}$ to be Hermitian. The matrix $\mat{A}_{CN}$ derived from the advection equation is generally non-Hermitian (due to the skew-symmetric first derivative). Since $\mat{A}$ is real-valued, we restore Hermiticity by left-multiplying by the transpose $\mat{A}^T$ and solving the resulting normal-equation system:
\begin{equation}
    \mat{A}^T \mat{A} \vct{x} = \mat{A}^T \vct{b}
    \label{eq:ata}
\end{equation}
The product $\mat{A}^T\mat{A}$ is symmetric positive definite when $\mat A$ is invertible. This is a normal-equation reformulation, not a preconditioner, and the right-hand side must also be transformed. It squares the condition number, $\kappa_2(\mat A^T\mat A)=\kappa_2(\mat A)^2$, which can be costly for poorly conditioned systems. Equation~\eqref{eq:conditioning} shows that the penalty is small for the present CN stencil, so the transformation alone does not establish a cause of hardware degradation.

For a complex nonsingular implicit system, the corresponding construction uses $\mat A^\dagger\mat A$ and $\mat A^\dagger\vct b$. A Hermitian block embedding is another option \cite{ref:hhl}; normal equations are not the only way to handle a non-Hermitian matrix. Applying either route to more general CFD equations requires suitable matrix access and preparation of the transformed input. Variable coefficients and nonperiodic boundaries also require reconsidering the Fourier implementation used here.

\subsection{Circulant HHL}\label{sec:circulant}
The constant-coefficient stencil and periodic boundary conditions make the matrices circulant. Their eigenvectors are Fourier modes. Writing $\mat A_h=\mat A^T\mat A$ and $\vct b_h=\mat A^T\vct b$ for CN, or $\mat A_h=\mat I$ and $\vct b_h=\vct b$ for the explicit schemes, the code uses
\begin{equation}
\mat A_h=\mat F\operatorname{diag}(\mu_k)\mat F^\dagger,
\qquad \mu_k=1+\frac{\alpha^2}{4}\sin^2\!\left(\frac{2\pi k}{N_c}\right)
\label{eq:fourier_spectrum}
\end{equation}
for CN; the explicit eigenvalues are all one. Here $\mat F$ is the Fourier matrix under the convention implemented in the code. The eigenvalues are computed classically by applying a fast Fourier transform (FFT) to the matrix's first column. The quantum basis changes use quantum Fourier transform (QFT) circuits. The classical FFT and the quantum QFT therefore play different roles.

Such use of circulant structure is established \cite{ref:zhou}. Here, we implement and evaluate this structure for the explicit and implicit advection systems. 
Our HHL implementation supports dense evolution and evolution constructed from Fourier basis changes and diagonal phase gates. In the shared-QFT HHL construction, one system-register QFT pair encloses the forward controlled evolution sequence, and another pair is used for uncomputation. The clock and reciprocal rotation remain present. This changes how matrix evolution is implemented within HHL; it does not replace HHL's phase-estimation and inversion stages. 

The known eigenvalues also permit a separate construction without phase estimation. Prepare $|b_h\rangle$, apply $\mat F^\dagger$, perform an ancilla rotation of angle $2\arcsin(C_\mu/\mu_k)$ controlled by Fourier index $k$, and apply $\mat F$. Here $0<C_\mu\le\min_k\mu_k$. Conditioning on ancilla outcome $1$ gives a state proportional to $\mat A_h^{-1}\vct b_h$. This clock-free Fourier multiplier is the construction used in all reported hardware fields and their noiseless finite-shot comparisons. It is distinct from full HHL, which is validated in statevector simulation.

\begin{figure}[!htbp]
\centering
\includegraphics[width=0.66\textwidth]{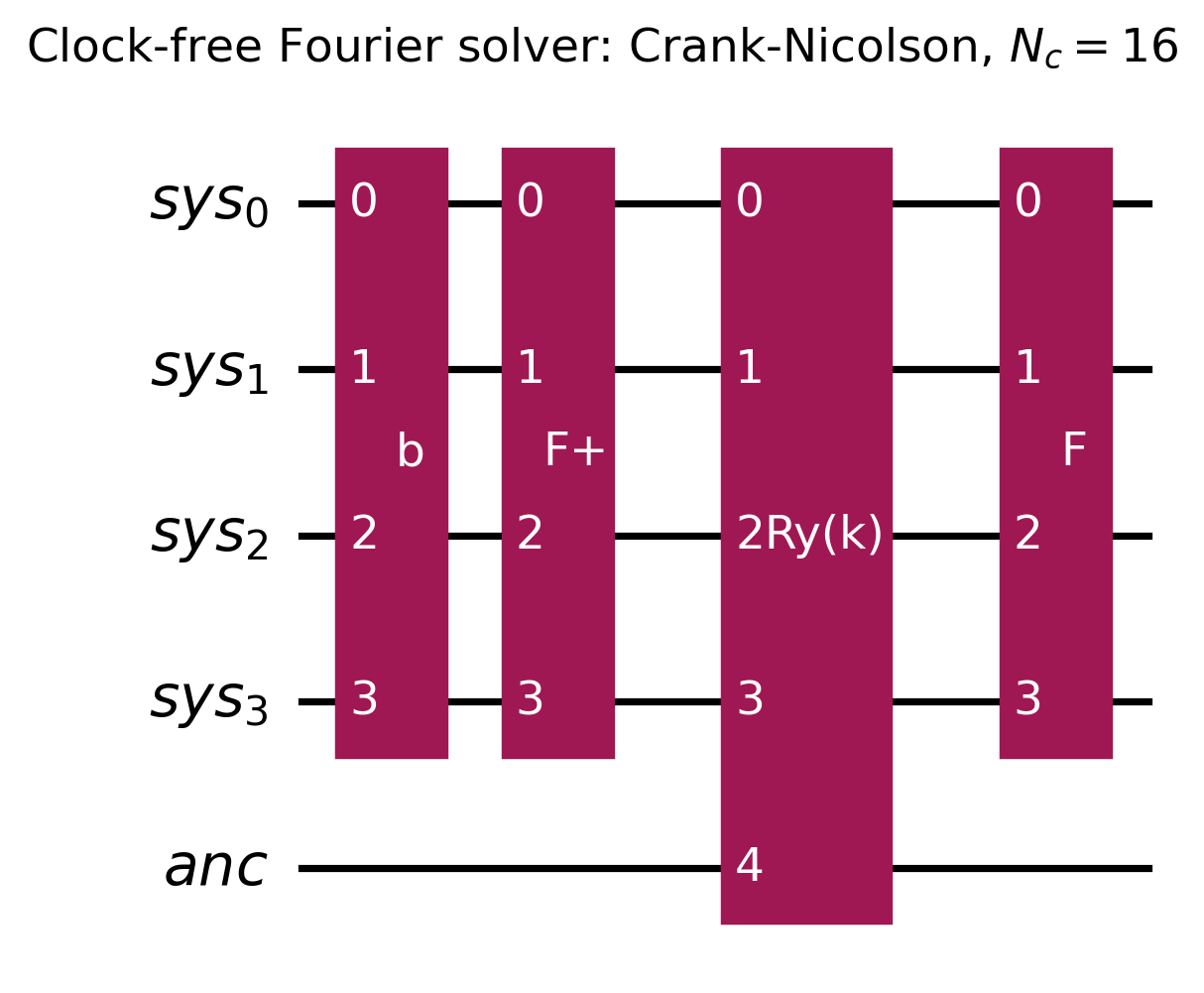}
\caption{Logical clock-free Fourier multiplier for CN at $N_c=16$. This construction is transpiled for hardware execution. The diagram omits measurements and physical routing; the reciprocal rotations use the classically known eigenvalues.}
\label{fig:fourier_circuit}
\end{figure}
\subsection{Circuit Implementation Details}
HHL uses a system register, a clock register, and one success ancilla. The system register has $n_{\rm sys}=\log_2N_c$ qubits. The comparison implementation uses three clock qubits for the explicit schemes and five ($N_c\le8$) or six ($N_c\ge16$) for CN. The Fourier multiplier uses only $n_{\rm sys}+1$ logical qubits. Its spectrum and controlled rotations are specified classically, and arbitrary amplitude preparation has a cost that grows with input size. We do not assume these operations are free in an end-to-end speedup claim. The software uses Qiskit~2.5.2, Aer~0.17.2, and IBM Runtime~0.49.0 \cite{ref:qiskit}.

\begin{figure}[!htbp]
    \centering
    \begin{subfigure}[b]{\textwidth}
        \centering
        \includegraphics[width=\textwidth]{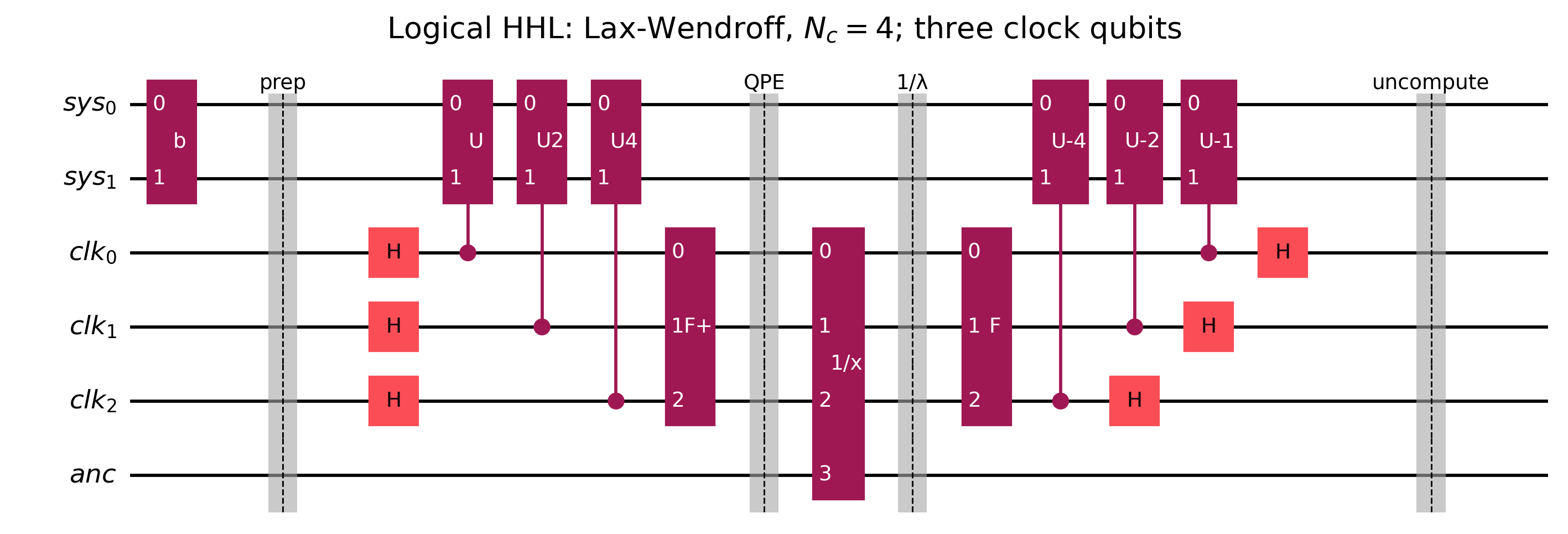}
        \caption{Logical HHL at $N_c=4$: two system qubits and three clock qubits}
    \end{subfigure}

    \vspace{1em}
    \begin{subfigure}[b]{\textwidth}
        \centering
        \includegraphics[width=\textwidth]{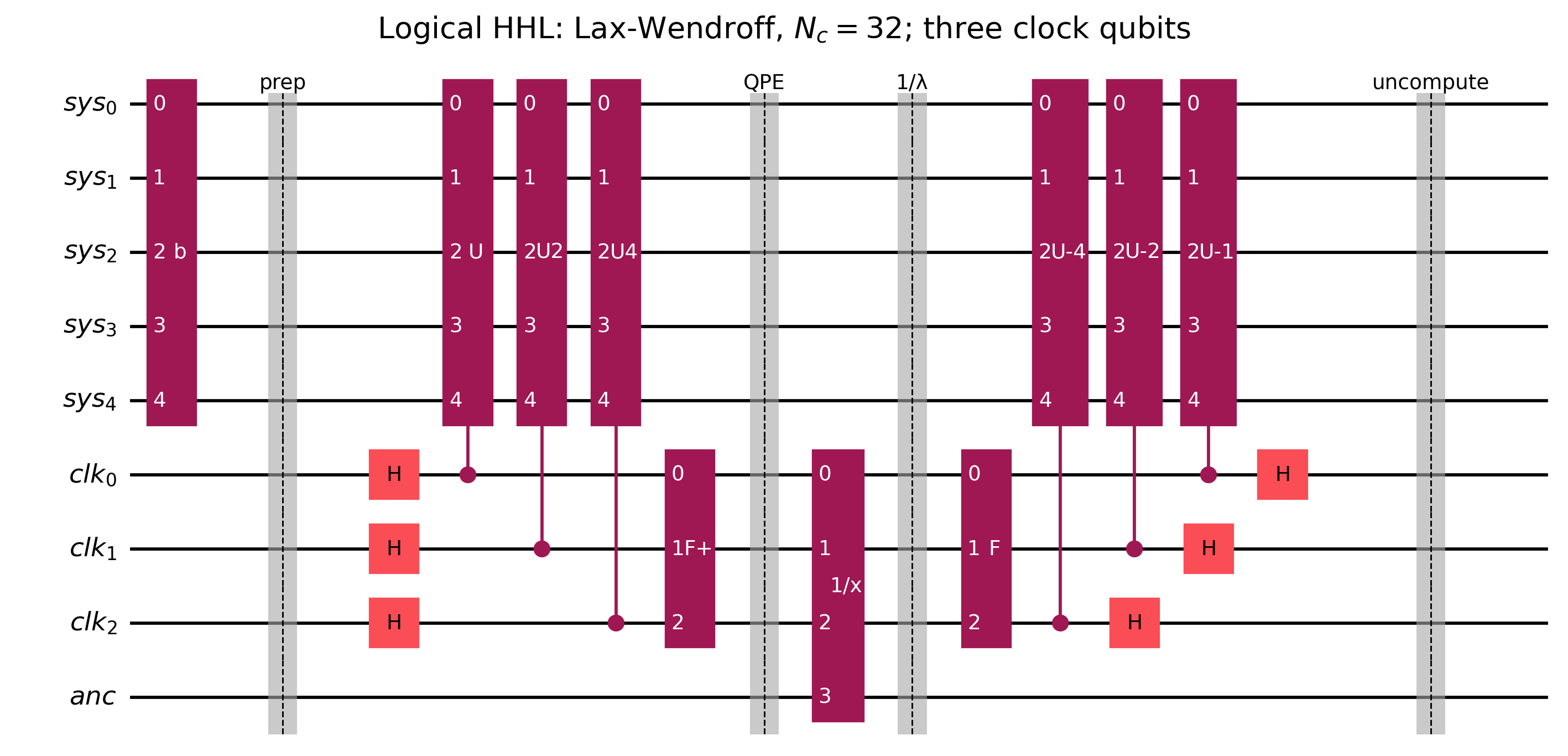}
        \caption{Logical HHL at $N_c=32$: five system qubits and three clock qubits}
    \end{subfigure}
    \caption{HHL circuits for the explicit LW system, showing preparation, phase estimation, reciprocal rotation, and uncomputation. The additional ancilla is used for postselection. Both diagrams illustrate noiseless HHL validation; the hardware uses the Fourier circuit in Fig.~\ref{fig:fourier_circuit}.}
    \label{fig:circuit_compare}
\end{figure}

Figure~\ref{fig:circuit_compare} illustrates the logical HHL implementation for the LW final-step system at $N_c=4$ and $32$. Each example has three clock qubits and uses $\tau=\pi$ to encode the identity eigenvalue exactly. The conditional state agrees with the direct solution to numerical precision. These are simulation illustrations, not transpiled hardware circuits.

Tables~\ref{tab:metrics_explicit} and~\ref{tab:metrics_implicit} report the resources of the actual submitted Fourier circuits in the \texttt{ibm\_rensselaer} native gate set, separately from the HHL illustrations.

\begin{table}[!htbp]
\centering
\caption{Submitted clock-free Fourier circuits for the explicit Lax--Wendroff scheme on \texttt{ibm\_rensselaer} (Nighthawk revision~1).}
\label{tab:metrics_explicit}
\small
\setlength{\tabcolsep}{5pt}
\begin{tabular}{@{}rrrrrrr@{}}
\toprule
$N_c$ & $n_{\rm sys}$ & $q_{\rm log}$ & $q_{\rm act}$ & Depth & CZ & Shots \\
\midrule
4 & 2 & 3 & 4 & 187 & 52 & 20,000 \\
8 & 3 & 4 & 4 & 810 & 222 & 20,000 \\
16 & 4 & 5 & 5 & 2,069 & 809 & 20,000 \\
32 & 5 & 6 & 8 & 5,557 & 2,324 & 40,000 \\
64 & 6 & 7 & 9 & 14,997 & 6,441 & 80,000 \\
\bottomrule
\end{tabular}
\par\smallskip
\begin{minipage}{0.97\textwidth}
\footnotesize $n_{\rm sys}=\log_2 N_c$ counts system qubits; $q_{\rm log}=n_{\rm sys}+1$ includes the success ancilla; $q_{\rm act}$ counts all physical qubits touched by native operations, including routing. Depth and CZ count describe the submitted computational-basis circuit. Each job also contains $\log_2 N_c$ Hadamard-basis circuits, with the same shots per circuit. The $N_c=4,8$ rows are resource records for fixed-Courant experiments ending at $t=2,1$; only $N_c=16,32,64$ match the field comparison at $t=0.5$. The LW $N_c=64$ job uses the August~25 calibration and a different layout; the other rows use the August~24 calibration.
\end{minipage}
\end{table}

\begin{table}[!htbp]
\centering
\caption{Submitted clock-free Fourier circuits for the implicit Crank--Nicolson normal equations on \texttt{ibm\_rensselaer} (Nighthawk revision~1).}
\label{tab:metrics_implicit}
\small
\setlength{\tabcolsep}{5pt}
\begin{tabular}{@{}rrrrrrr@{}}
\toprule
$N_c$ & $n_{\rm sys}$ & $q_{\rm log}$ & $q_{\rm act}$ & Depth & CZ & Shots \\
\midrule
4 & 2 & 3 & 4 & 186 & 52 & 20,000 \\
8 & 3 & 4 & 4 & 799 & 222 & 20,000 \\
16 & 4 & 5 & 5 & 2,066 & 809 & 20,000 \\
32 & 5 & 6 & 8 & 5,568 & 2,324 & 40,000 \\
64 & 6 & 7 & 8 & 17,436 & 6,860 & 80,000 \\
\bottomrule
\end{tabular}
\par\smallskip
\begin{minipage}{0.97\textwidth}
\footnotesize $n_{\rm sys}=\log_2 N_c$ counts system qubits; $q_{\rm log}=n_{\rm sys}+1$ includes the success ancilla; $q_{\rm act}$ counts all physical qubits touched by native operations, including routing. Depth and CZ count describe the submitted computational-basis circuit. Each job also contains $\log_2 N_c$ Hadamard-basis circuits, with the same shots per circuit. The $N_c=4,8$ rows are resource records for fixed-Courant experiments ending at $t=2,1$; only $N_c=16,32,64$ match the field comparison at $t=0.5$.
\end{minipage}
\end{table}

\subsection{Hardware and Error Metrics}
We use Qiskit \texttt{AerSimulator} for noiseless sampling and the IBM \texttt{ibm\_rensselaer} backend at Rensselaer Polytechnic Institute. Archived device metadata identify a 120-qubit Nighthawk revision~1 processor with native CZ gates. The submitted circuits use 4--9 active physical qubits after routing. Nineteen of the twenty scheme--grid hardware jobs use the August~24, 2026 calibration; LW at $N_c=64$ uses the August~25 calibration and a different layout. Job identifiers, native circuits, and their associated calibration records are retained with the data.
The field-error plots use the exact translated packet $\vct\phi_{\rm exact}$ evaluated on the 128-point grid. If $\widehat{\vct x}_c$ is the reconstructed coarse field and $\mathcal I_{\rm per}$ denotes periodic linear interpolation, the relative error is
\begin{equation}
E=\frac{\|\mathcal I_{\rm per}\widehat{\vct x}_c-\vct\phi_{\rm exact}\|_2}{\|\vct\phi_{\rm exact}\|_2}.
\label{eq:field_error}
\end{equation}
The fine-grid classical solution is a separate visual reference. To isolate reconstruction from discretization and interpolation, we also use the native error
\begin{equation}
e=\frac{\|\widehat{\vct x}_c-\vct x_c\|_2}{\|\vct x_c\|_2},
\label{eq:native_error}
\end{equation}
where $\vct x_c$ is the exact classical solution of the same coarse discrete system. The same definitions are applied separately to noiseless finite-shot and hardware reconstructions. These error norms are not additive.

Each measurement setting uses 20,000 shots for $N_c\le16$, 40,000 for $32$, and 80,000 for $64$. The hardware data contain twenty scheme--grid jobs and 100 measurement settings. Only the twelve jobs at $N_c=16,32,64$ match the common final time $T_s=0.5$: their matrices and right-hand sides agree with the common-time systems. The $N_c=4,8$ fixed-Courant hardware runs reached $t=2,1$ and are included only as resource records, not as fields at $T_s=0.5$. No measurement-error mitigation is applied. Circuit depth is reported as a gate-layer count; it cannot be compared directly with a coherence time without circuit-duration information.

A signed field includes both positive and negative values. Computational-basis measurements conditioned on ancilla $1$ determine the magnitudes of the normalized solution. An additional Hadamard-basis setting for each system qubit gives interference information for relative-sign recovery, for a total of $\log_2N_c+1$ settings. The reconstruction propagates these signs along a maximum spanning tree weighted toward larger amplitudes. Small amplitudes and disconnected support can make sign recovery unstable or ambiguous.

The field norm is estimated from the computational-basis ancilla success fraction $\widehat P_{\rm comp}$ using
\begin{equation}
\widehat{\|\vct x_c\|_2}=\frac{\sqrt{\widehat P_{\rm comp}}\,\|\vct b_h\|_2}{C_\mu}.
\label{eq:field_norm}
\end{equation}
This relation is exact for the ideal Fourier multiplier, but hardware noise can bias the estimate. The reconstructed field is aligned with the exact coarse solution by one global sign before computing errors; component magnitudes and relative signs otherwise come from the counts. Ancilla success alone is not a measure of solution fidelity.

\section{Results and Analysis}\label{sec:results}
In this section we compare the classical and quantum results. We first establish the classical baseline for the four finite difference schemes on the fine grid. We then compare noiseless finite-shot Fourier reconstructions with execution of the same construction on \texttt{ibm\_rensselaer}. Finally, we examine field error and circuit resources as functions of coarse-grid resolution, retaining the distinction between explicit controls and the implicit CN system.

\subsection{Classical Benchmarks}
\begin{figure}[!htbp]
  \centering
  \includegraphics[width=\textwidth]{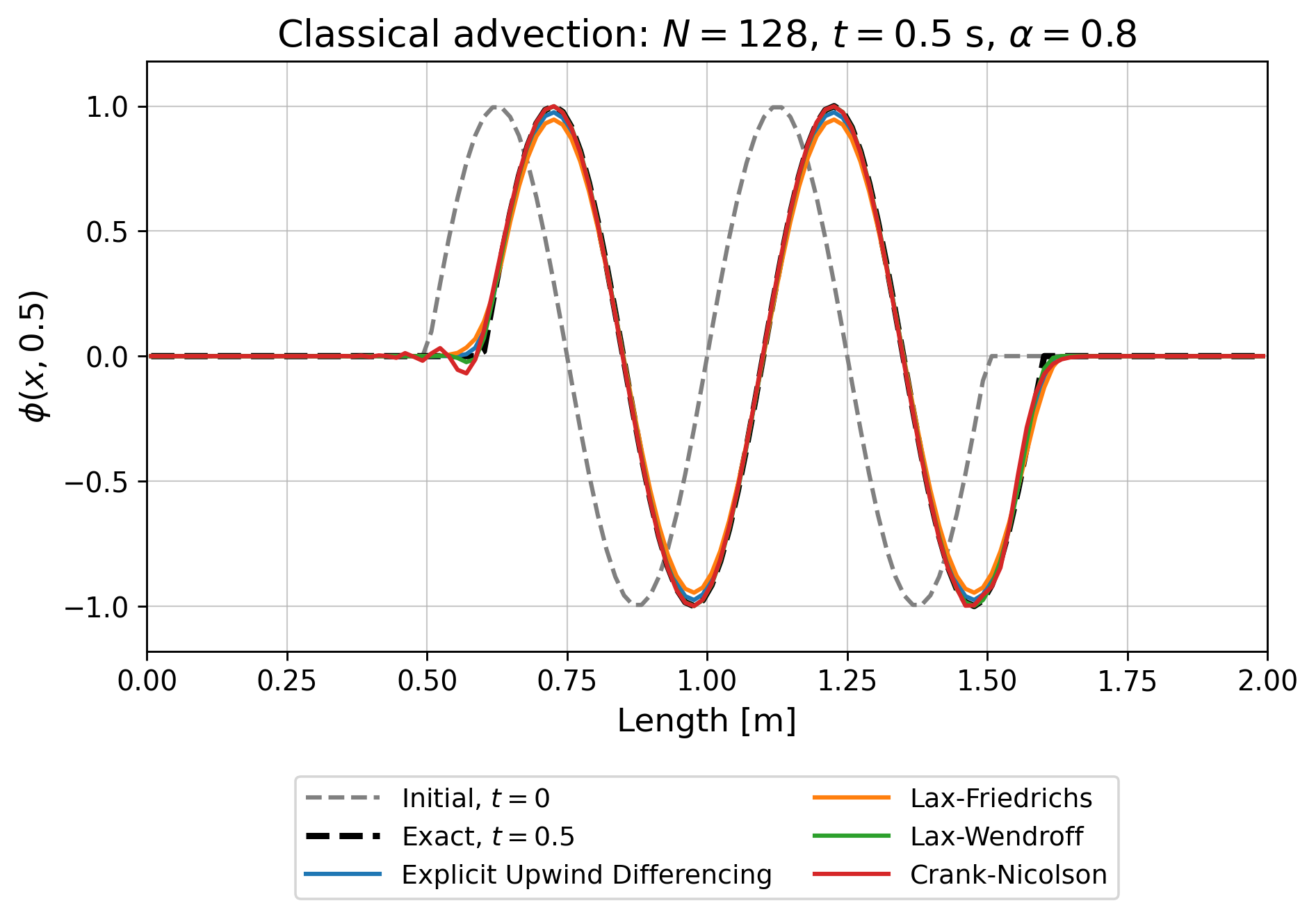}
  \caption{Classical FD schemes on the fine grid ($N=128$) at $T_s=0.5$. The gray curve is the initial condition, the black dashed curve is its exact periodic translation, and the colored curves show the four numerical schemes.}
  \label{fig:propagating_wave}
\end{figure}

Figure~\ref{fig:propagating_wave} illustrates the behavior of the classical schemes. LW and CN retain more of the packet amplitude but introduce dispersive error, whereas the more dissipative EUD and LF schemes damp the peak amplitude. The exact translated packet provides a reference for these numerical differences.
\clearpage

\subsection{Quantum Simulation vs. Hardware Execution}
HHL statevector tests validate the discrete linear-solver implementation, while the profiles below evaluate signed Fourier reconstructions with finite measurements. All displayed fields reach $T_s=0.5$. The black solid curve is the exact transported packet and the gray dotted curve is the fine-grid classical solution for the corresponding scheme. Hardware profiles are available at $N_c=16,32,64$. The observed discrepancies include circuit and readout effects and sensitivity of sign and norm reconstruction; these data alone do not identify a universal hardware failure threshold.

\subsubsection{Explicit Schemes (EUD, LF, LW)}
\begin{figure}[!htbp]
    \centering
    \begin{subfigure}[b]{\textwidth}
        \centering
        \includegraphics[width=\textwidth]{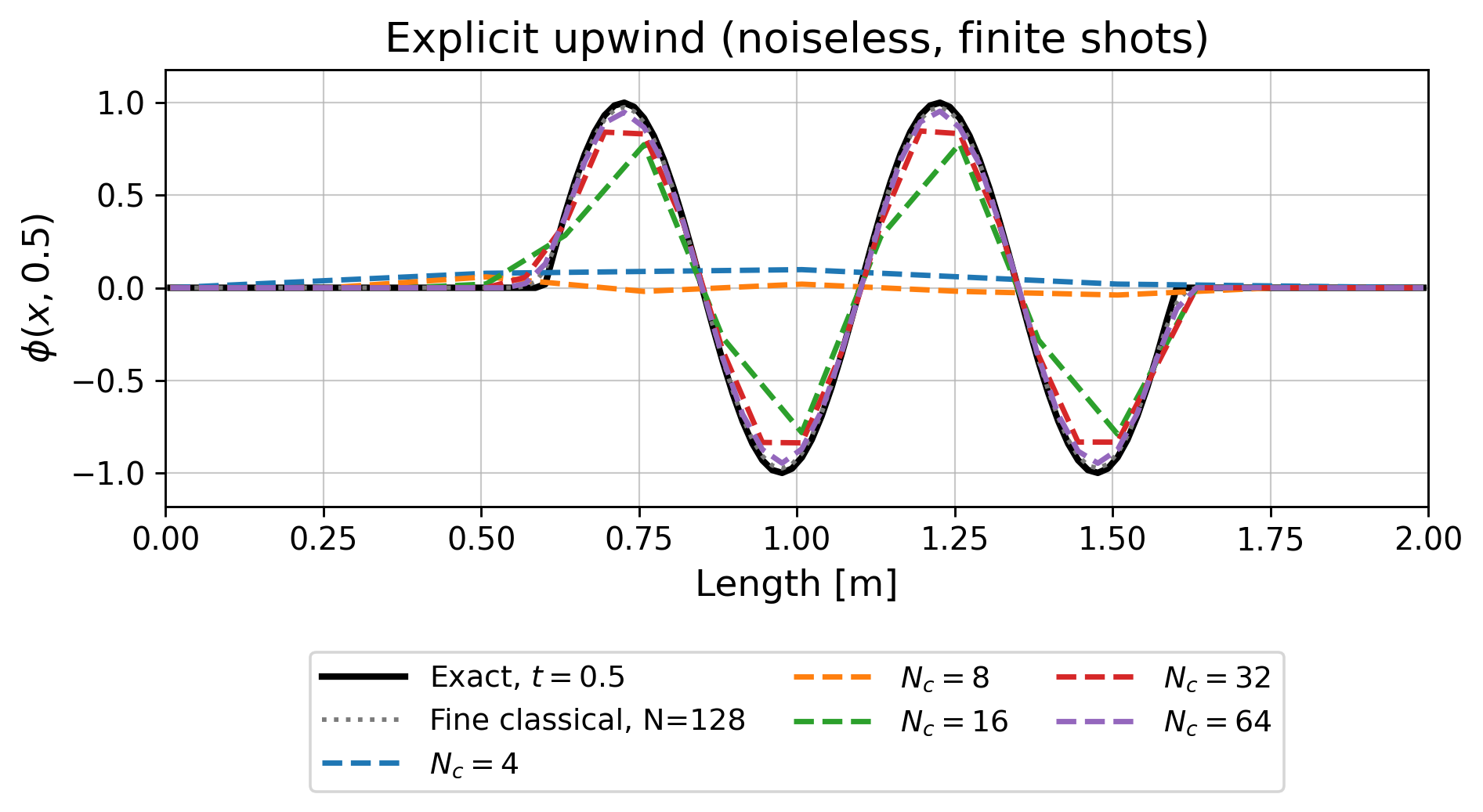}
        \caption{Noiseless finite-shot Fourier reconstruction (Explicit Upwind)}
    \end{subfigure}
    \vspace{1em}
    \begin{subfigure}[b]{\textwidth}
        \centering
        \includegraphics[width=\textwidth]{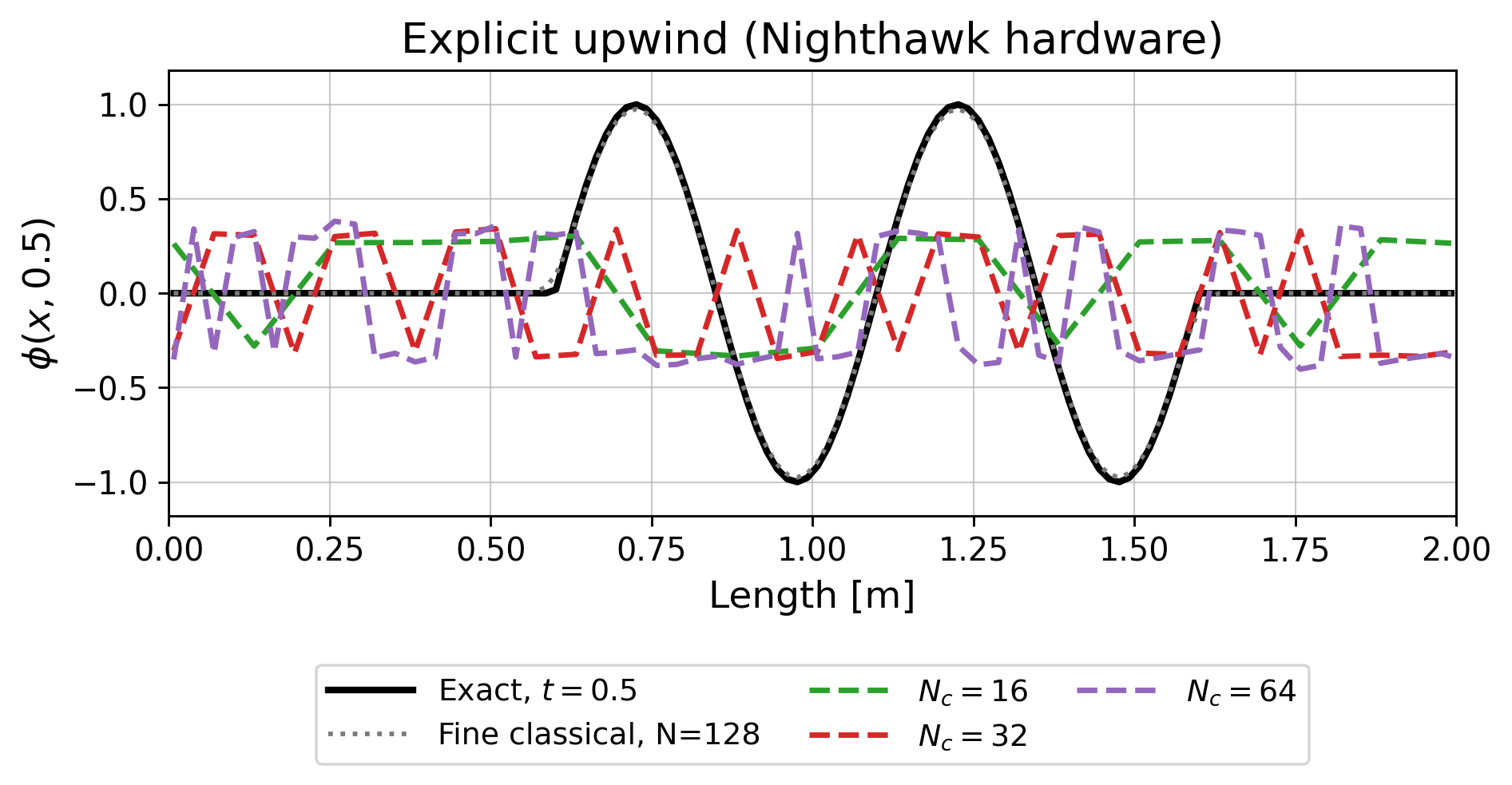}
        \caption{Hardware Execution (Explicit Upwind)}
    \end{subfigure}
    \caption{Explicit Upwind comparison at $T_s=0.5$. The exact transported packet and fine-grid classical solution are shown as references. Hardware curves include only $N_c=16,32,64$.}
    \label{fig:eud_compare}
\end{figure}

\begin{figure}[!htbp]
    \centering
    \begin{subfigure}[b]{\textwidth}
        \centering
        \includegraphics[width=\textwidth]{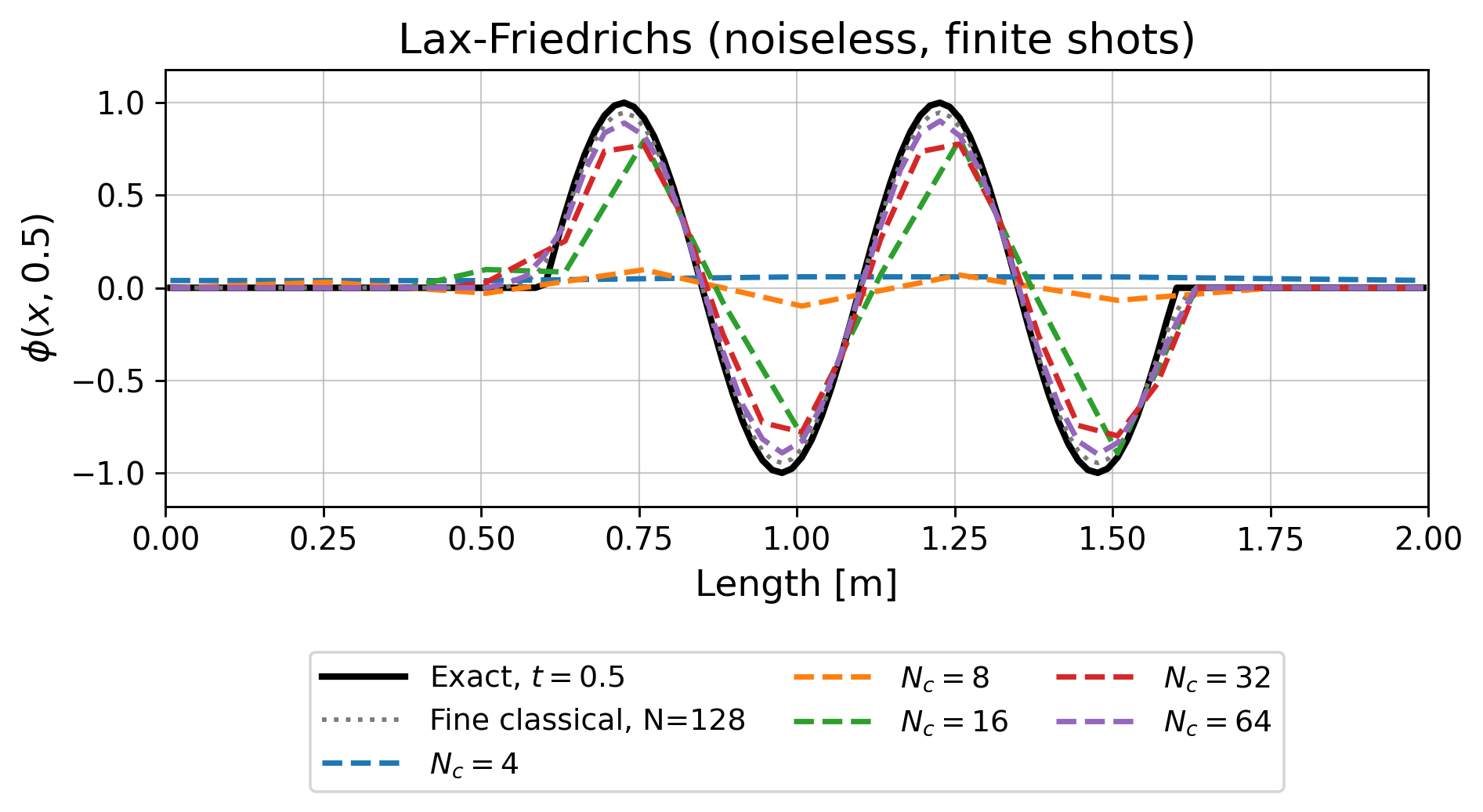}
        \caption{Noiseless finite-shot Fourier reconstruction (Lax--Friedrichs)}
    \end{subfigure}
    \vspace{1em}
    \begin{subfigure}[b]{\textwidth}
        \centering
        \includegraphics[width=\textwidth]{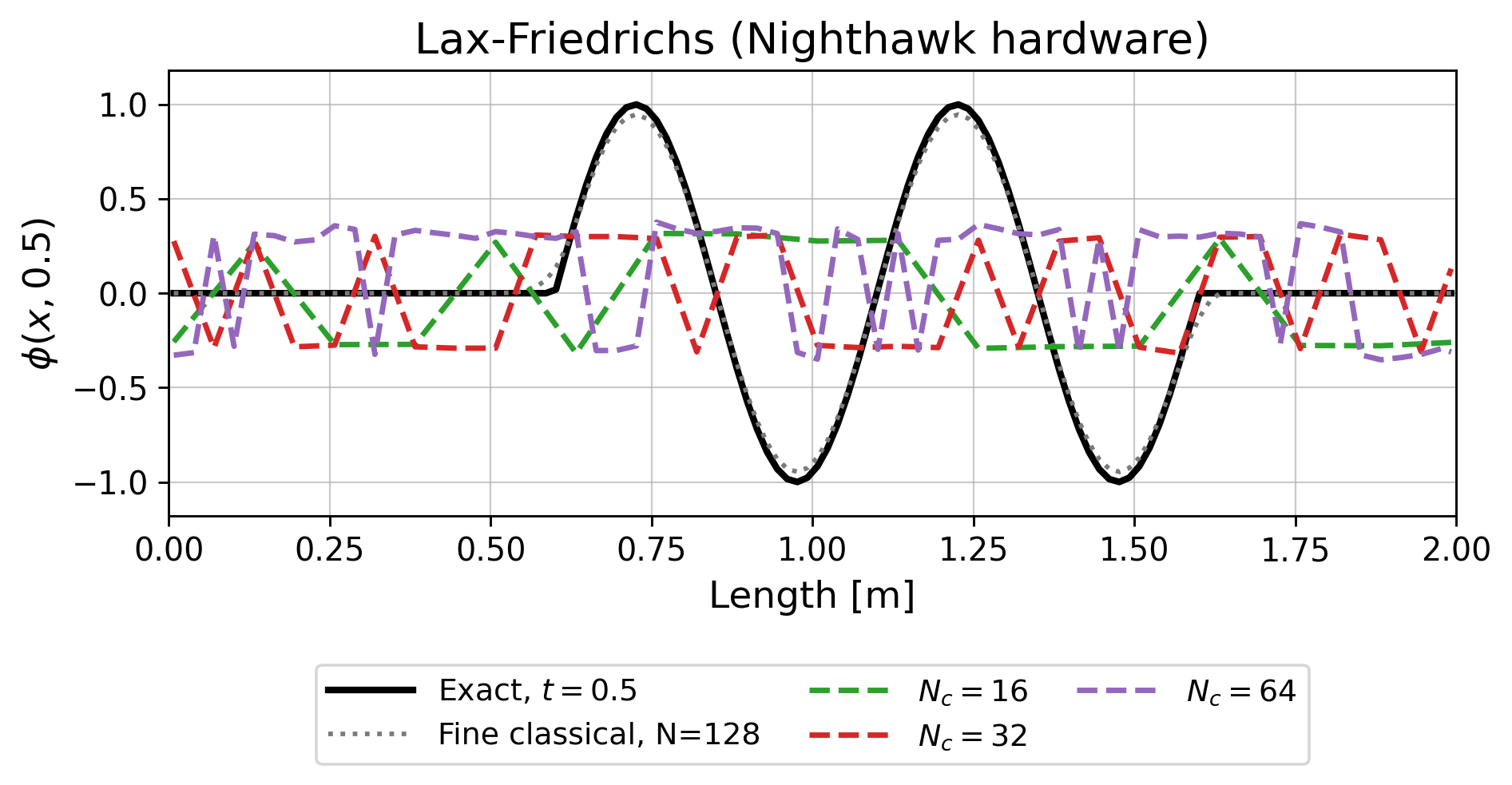}
        \caption{Hardware Execution (Lax--Friedrichs)}
    \end{subfigure}
    \caption{Lax--Friedrichs comparison at $T_s=0.5$. The noiseless curves retain the numerical damping of the scheme; hardware introduces additional reconstruction error. Only $N_c=16,32,64$ have hardware data for this final time.}
    \label{fig:lf_compare}
\end{figure}

\begin{figure}[!htbp]
    \centering
    \begin{subfigure}[b]{\textwidth}
        \centering
        \includegraphics[width=\textwidth]{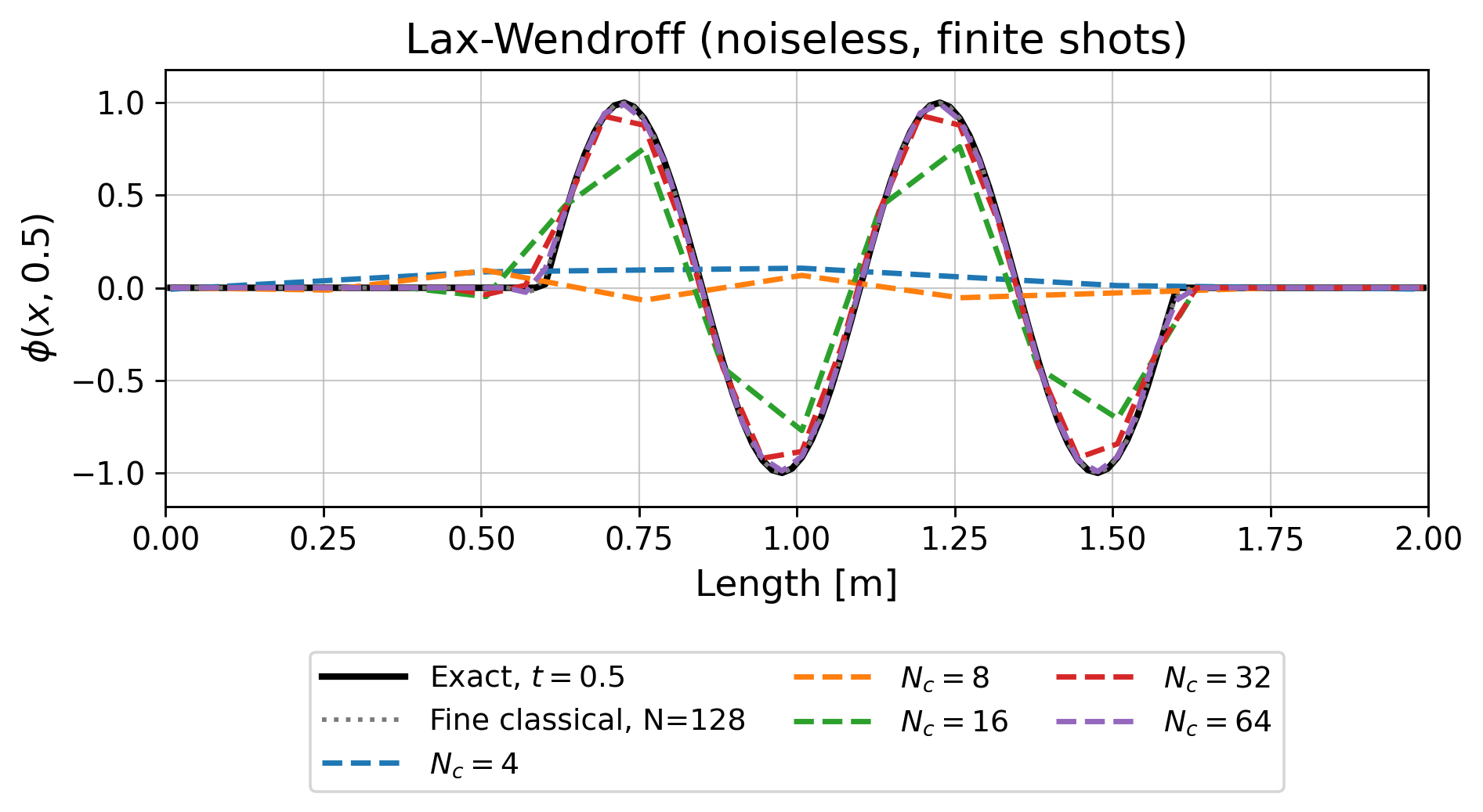}
        \caption{Noiseless finite-shot Fourier reconstruction (Lax--Wendroff)}
        \label{fig:lw_sim}
    \end{subfigure}
    \vspace{1em}
    \begin{subfigure}[b]{\textwidth}
        \centering
        \includegraphics[width=\textwidth]{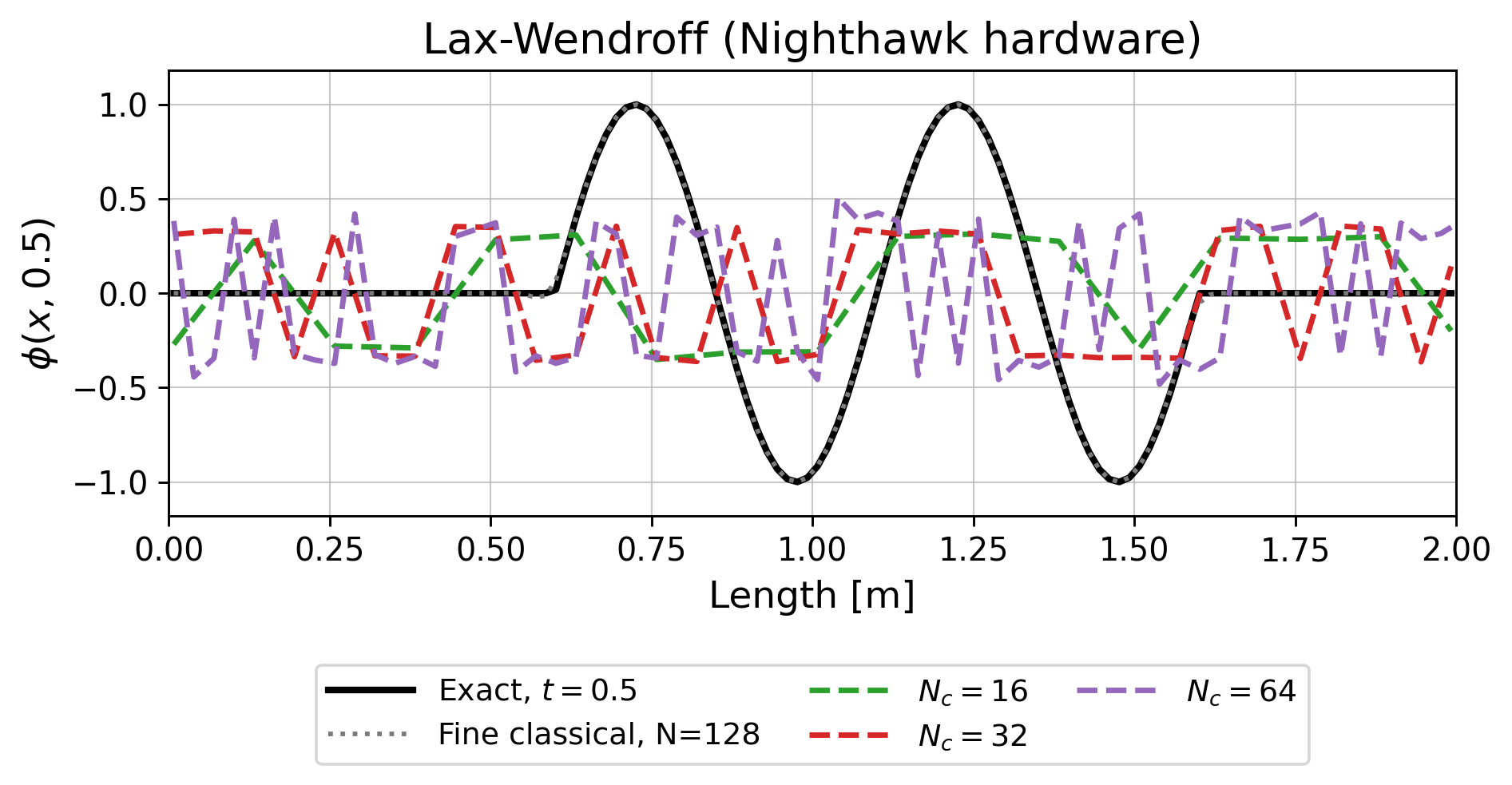}
        \caption{Hardware Execution (Lax--Wendroff)}
        \label{fig:lw_real}
    \end{subfigure}
    \caption{Lax--Wendroff comparison at $T_s=0.5$. Hardware reconstruction errors occur across the measured grids $N_c=16,32,64$; the curves do not establish a universal onset of failure at 32 points.}
    \label{fig:lw_compare}
\end{figure}

For the explicit schemes, the coefficient matrix is the identity (Eqs.~\ref{eq:eud_matrix}, \ref{eq:lf_matrix}, and~\ref{eq:lw_matrix}). Their right-hand sides differ because EUD, LF, and LW have different dissipation and dispersion. Figures~\ref{fig:eud_compare}, \ref{fig:lf_compare}, and~\ref{fig:lw_compare} show that the noiseless reconstructions approach the exact transported packet on the resolved grids, while the $N_c=4,8$ samples poorly resolve it. At $N_c=64$, the noiseless field errors are 0.069, 0.127, and 0.032 for EUD, LF, and LW. The hardware errors are 1.165, 1.143, and 1.186, respectively, and substantial errors already occur at $N_c=16$. The data do not support a claim that all three schemes remain accurate up to a fixed 32-point threshold.

\clearpage 

\subsubsection{Implicit Scheme (Crank-Nicolson)}
The implicit CN scheme (Fig.~\ref{fig:cn_compare}) requires the normal-equation solve in Eq.~\ref{eq:ata}. Although this squares the condition number, Eq.~\eqref{eq:conditioning} gives only $1.16$ for $\mat A^T\mat A$ at $\alpha=0.8$. The noiseless Fourier field errors decrease from 0.528 at $N_c=16$ to 0.200 at $32$ and 0.077 at $64$. The corresponding exact classical coarse-field errors are 0.530, 0.206, and 0.079, showing that the observed noiseless reconstructions largely follow the discretization and interpolation error in these samples.

The hardware CN errors are 0.665, 1.052, and 1.168 at $N_c=16,32,64$. CN has the smallest observed field error among the four schemes at 16 points, but it is not consistently best at the larger sizes. Its native reconstruction errors are 0.653, 1.106, and 1.219, compared with noiseless native errors of approximately 0.014, 0.014, and 0.011. Thus the hardware degradation is not evidence of a growing condition number or a phase-estimation error floor: the spectrum remains mild, and these Fourier circuits have no clock register. Each reported hardware case is one archived job, so these comparisons do not establish a repeatable advantage for CN.

\begin{figure}[!htbp]
    \centering
    \begin{subfigure}[b]{\textwidth}
        \centering
        \includegraphics[width=\textwidth]{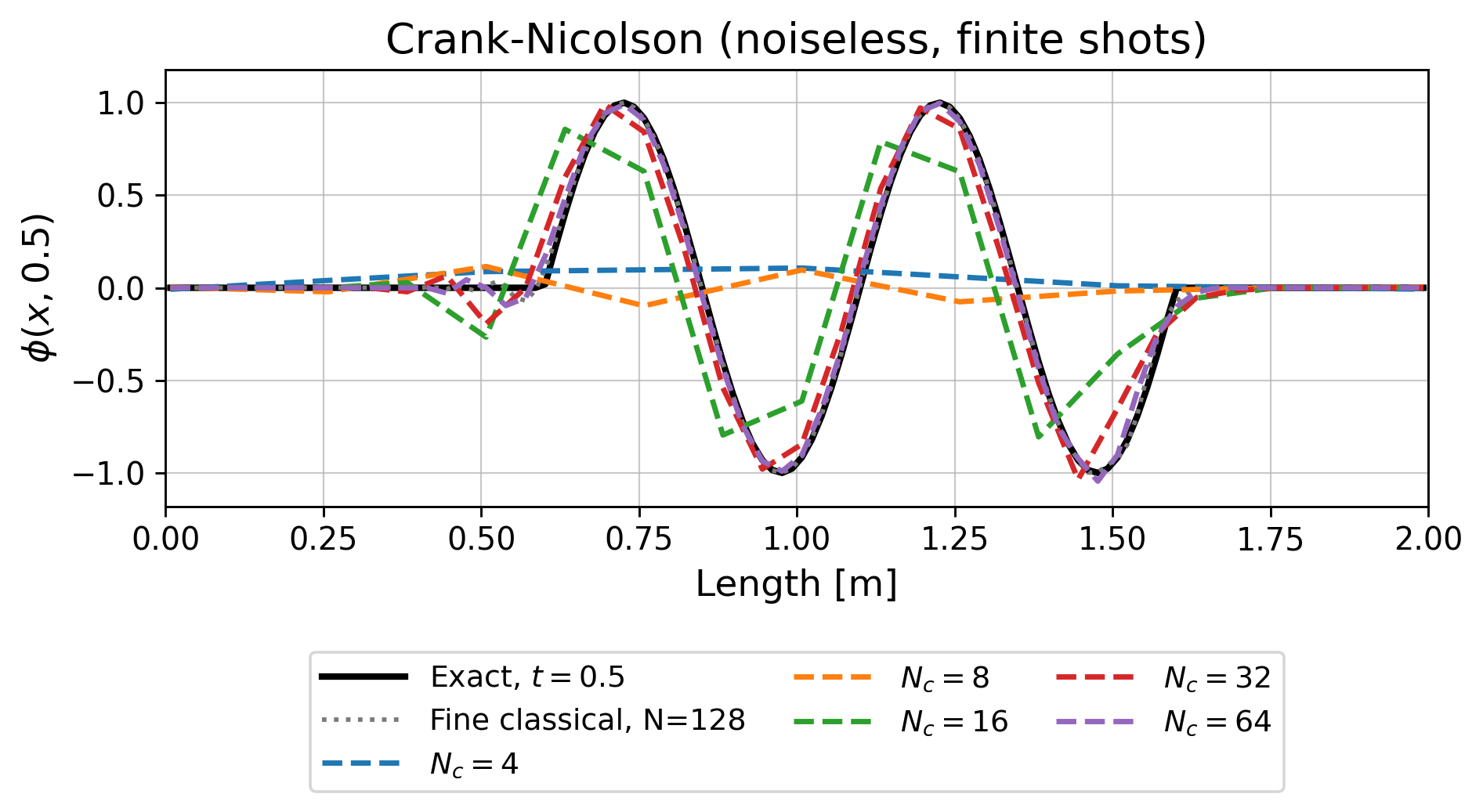}
        \caption{Noiseless finite-shot Fourier reconstruction (Crank--Nicolson)}
        \label{fig:cn_sim}
    \end{subfigure}
    \caption{Crank--Nicolson scheme comparison. (Figure continues on next page)}
    \label{fig:cn_compare} 
\end{figure}

\begin{figure}[H] 
    \ContinuedFloat 
    \centering
    \begin{subfigure}[b]{\textwidth}
        \centering
        \setcounter{subfigure}{1} 
        \includegraphics[width=\textwidth]{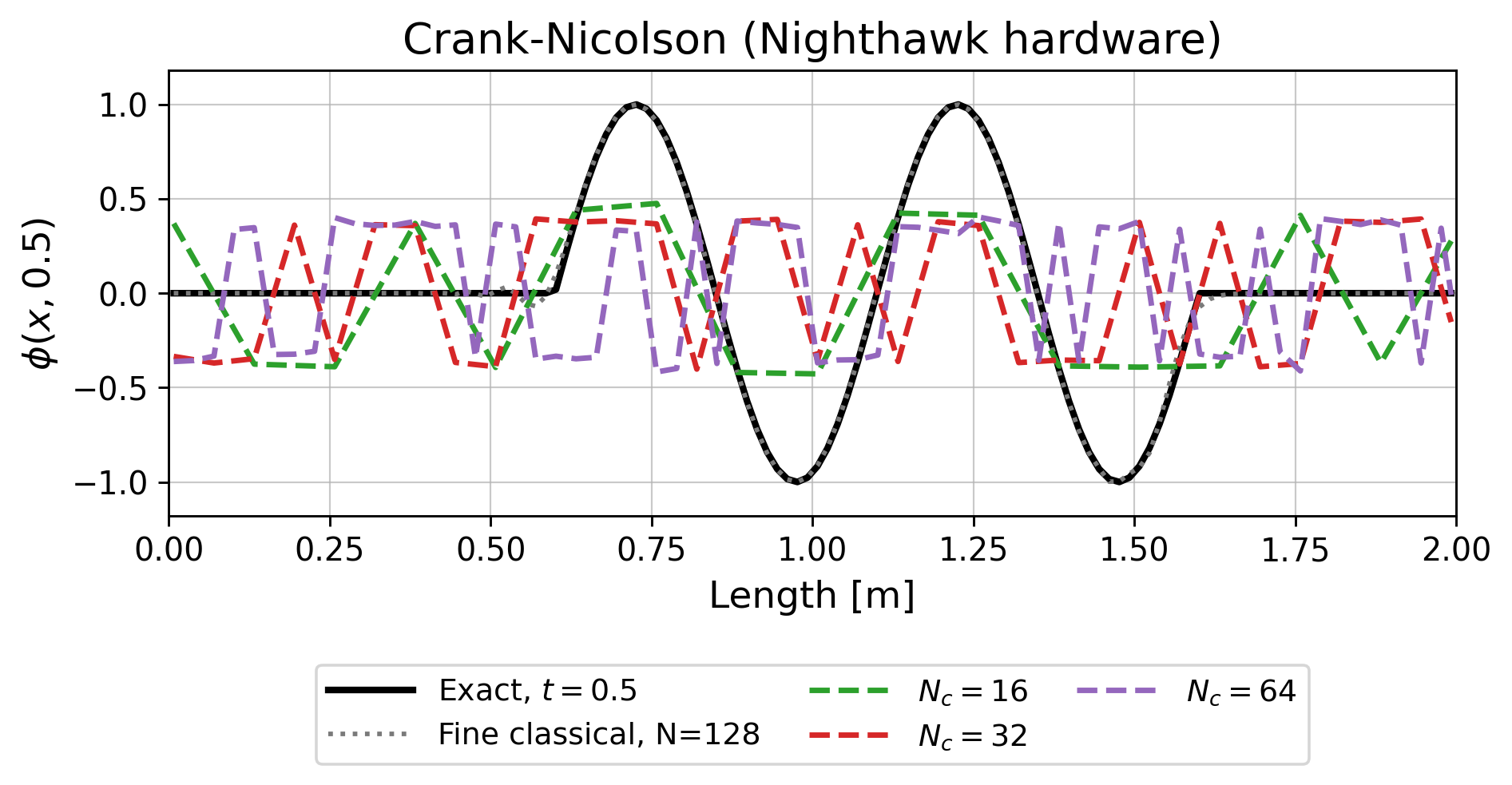}
        \caption{Hardware Execution (Crank--Nicolson)}
        \label{fig:cn_real}
    \end{subfigure}
    \caption{Crank--Nicolson comparison at $T_s=0.5$ (continued). The hardware Fourier reconstructions use the mildly conditioned normal-equation system. Their field errors cannot be attributed to the normal-equation condition number alone.}
\end{figure}

\subsection{Error Scaling Analysis}
Figure~\ref{fig:error_compare} summarizes errors against the exact transported packet at $T_s=0.5$. The noiseless curves decrease on the resolved grids; the hardware curves do not show the same improvement. No common-time hardware errors are reported at $N_c=4,8$. The observed trends do not establish an asymptotic convergence order or a universal failure threshold at $N_c=32$.

Two comparisons help interpret these results. The field error $E$ includes discretization, periodic interpolation, and reconstruction, whereas the native error $e$ compares the measured field with the exact discrete solution before interpolation. For the recorded noiseless reconstructions at $N_c=16,32,64$, native errors are approximately 0.008--0.014 across the four schemes. Finite sampling can still change relative-sign decisions, especially near zero amplitudes. In the accompanying resampling analysis, noiseless EUD at $N_c=16$ has observed field error 0.347 but a conditional 2.5--97.5 percentile range of approximately 0.346--0.934. These ranges describe sensitivity to the existing counts, not independent hardware repetitions or confidence intervals with established coverage. Good agreement of an individual noiseless sample is therefore not a guarantee of stable full-field recovery.

\begin{figure}[!htbp]
    \centering
    \begin{subfigure}[b]{\textwidth}
        \centering
        \includegraphics[width=\textwidth]{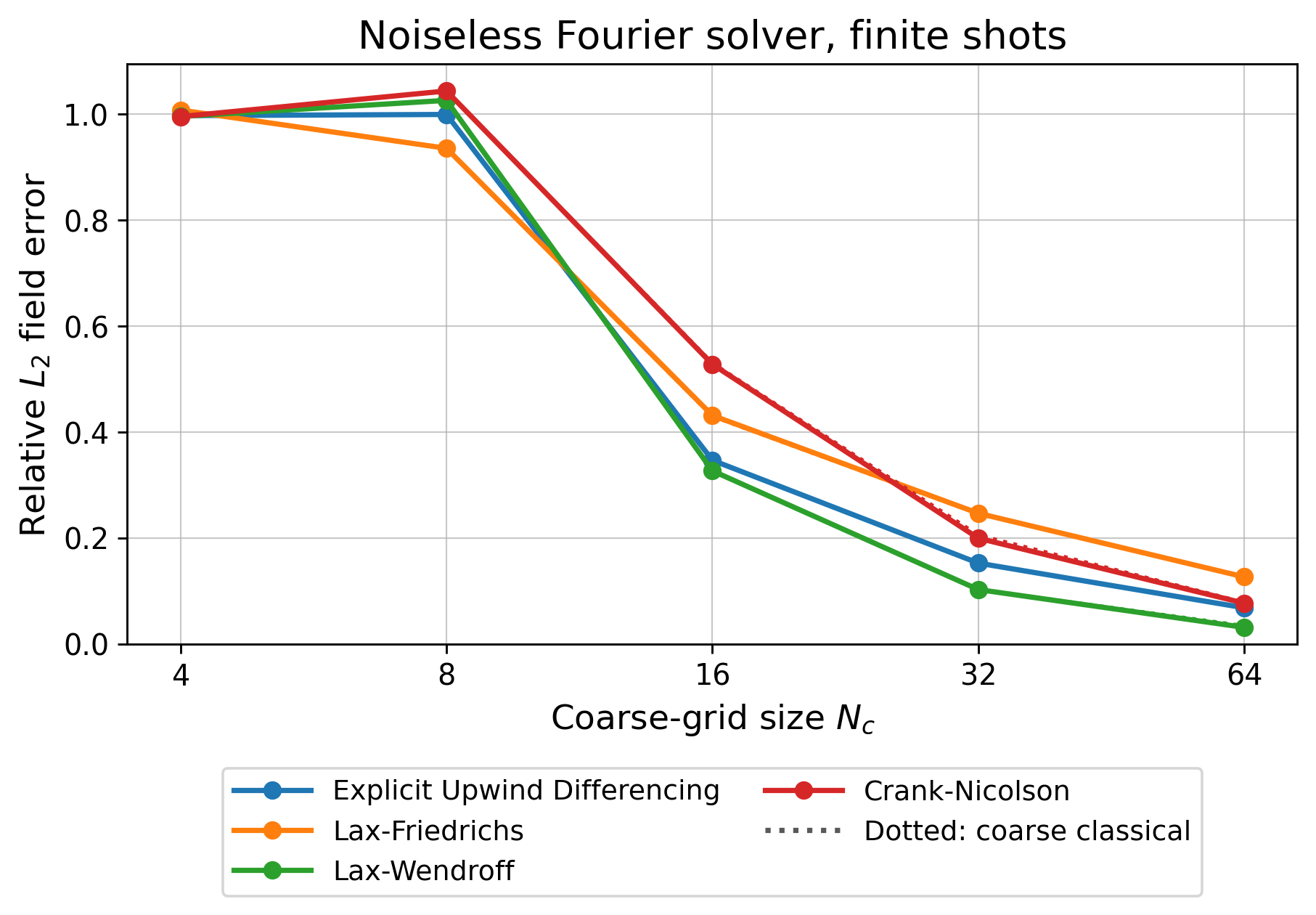}
        \caption{Noiseless finite-shot relative field error against the exact translated packet}
        \label{fig:error_sim}
    \end{subfigure}
\end{figure}

 \begin{figure}[!htbp]
    \ContinuedFloat 
    \centering
    \begin{subfigure}[b]{\textwidth}
        \centering
        \setcounter{subfigure}{1}
        \includegraphics[width=\textwidth]{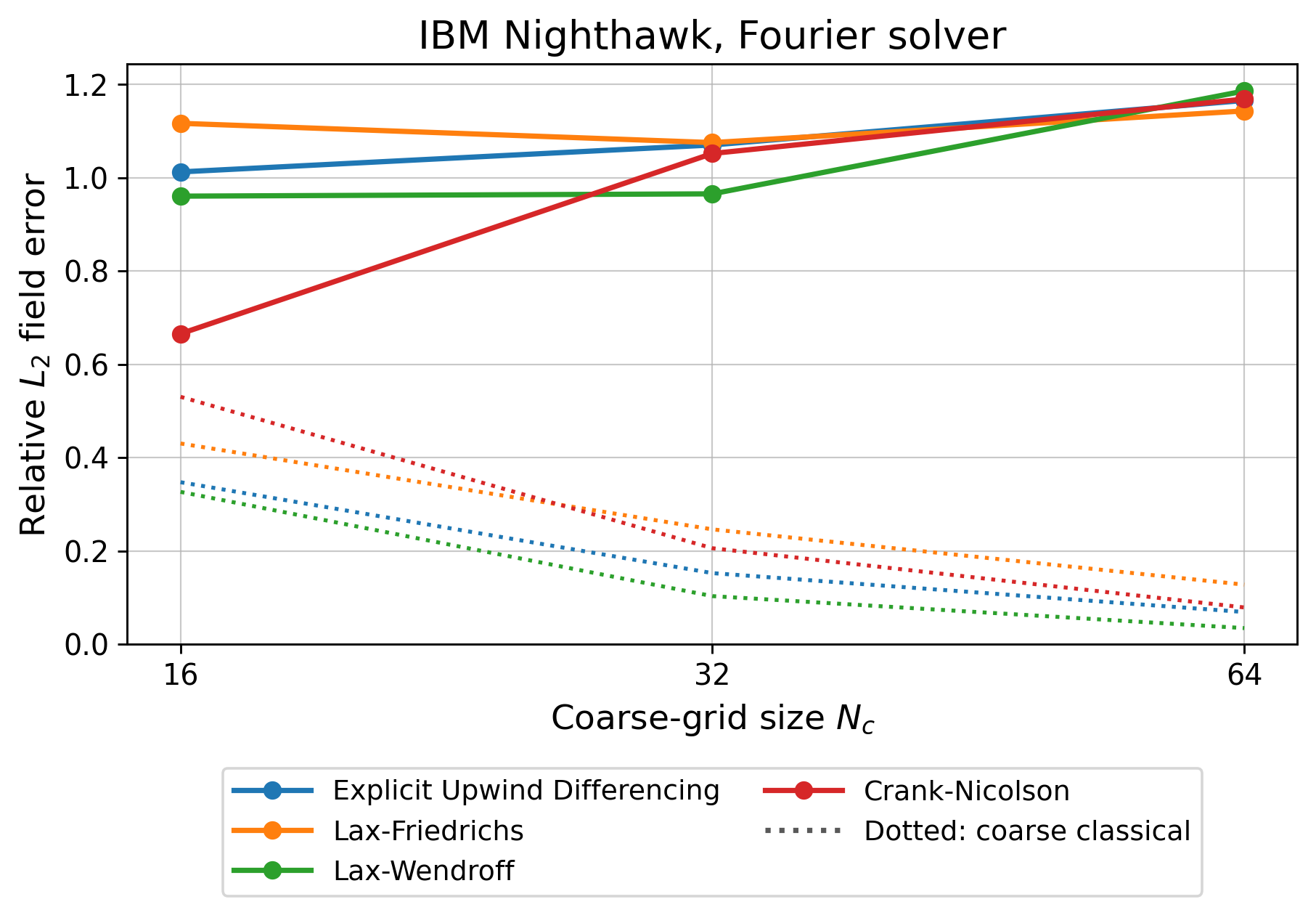}
        \caption{Hardware relative field error against the exact translated packet}
        \label{fig:error_real}
    \end{subfigure}
    \caption{Relative field error versus $N_c$ at $T_s=0.5$, using periodic interpolation. Dotted curves show the exact classical coarse-grid error under the same policy. Hardware points are limited to $N_c=16,32,64$. The smallest two grids in (a) poorly resolve the packet.}
    \label{fig:error_compare}
\end{figure}
\begin{figure}[!htbp]
\centering
\includegraphics[width=\textwidth]{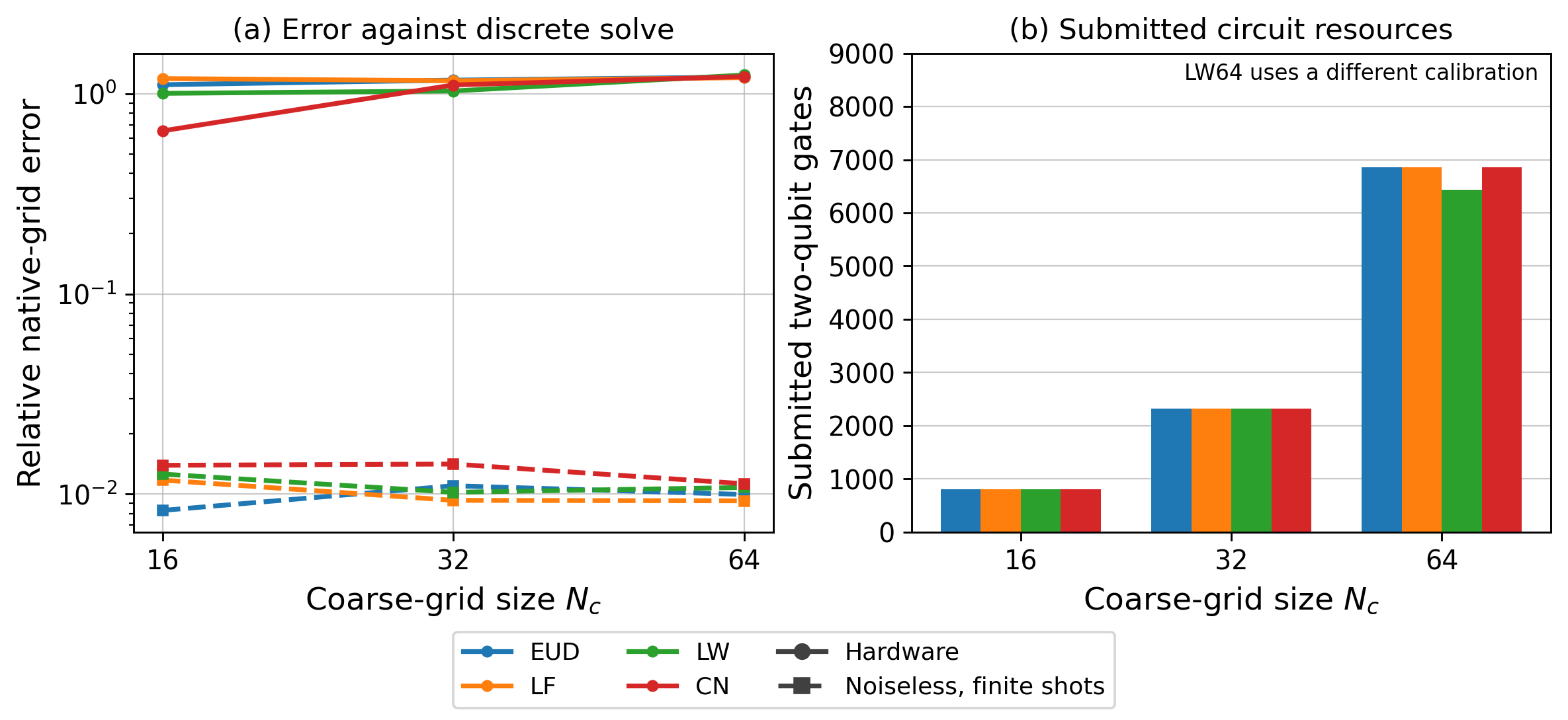}
\caption{CN versus the three explicit controls at $T_s=0.5$. Left: native error $e$ against the exact discrete solution, with dashed noiseless and solid hardware curves. Right: actual CZ counts for the submitted computational-basis circuits. LW at $N_c=64$ uses a different calibration and layout.}
\label{fig:cn_explicit}
\end{figure}

Figure~\ref{fig:cn_explicit} directly compares the native reconstruction errors and submitted CZ counts. At $N_c=16,32$, all four schemes use 809 and 2,324 CZ gates, respectively. CN depths are 2,066 and 5,568, close to LW's 2,069 and 5,557. At $N_c=64$, EUD, LF, and CN use 6,860 CZ gates; LW uses 6,441 but has a different calibration and layout. The explicit matrices are better conditioned than CN, but this small conditioning difference does not rank either the observed errors or the two-qubit counts in a simple way.

\clearpage
\section{Discussion}\label{sec:discussion}
The results illustrate the practical costs of applying quantum linear solvers to CFD. Theoretical HHL advantages depend on matrix access, conditioning, input preparation, and output requirements. In the present implementation, encoding classical data and reconstructing a signed field are explicit costs. Full HHL is assessed in noiseless statevector simulation; the hardware uses the clock-free Fourier construction, so hardware errors should not be attributed to QPE. This benchmark treats only the final time-step system quantum mechanically and does not demonstrate an end-to-end computational speedup.

For the explicit schemes (EUD, LF, LW), the linear system is trivial ($\mat A=\mat I$), and classical time-stepping is efficient. These schemes are controls for different prepared fields rather than competitive applications of a quantum inverse. Their different dissipation, dispersion, and small-amplitude components test reconstruction under three right-hand sides. They help assess input-dependent effects, but do not perfectly isolate inversion error from hardware or measurement error. All three controls exhibit substantial hardware field errors despite ideal matrix conditioning.

For CN, the normal equations provide a positive-definite system compatible with the implemented HHL reciprocal. This reformulation is one choice, and its condition number remains mild for the present stencil. CN's measured circuit resources are close to those of the explicit controls at 16 and 32 points, and its errors are not uniformly larger. A causal test of conditioning would require varying the spectrum while controlling inputs, compilation, layout, calibration, and readout. Future work may compare preconditioned formulations \cite{ref:q_pde_review}, variational quantum linear solvers \cite{ref:vqls}, or other implementations for non-Hermitian CFD systems, but their practical accuracy and cost would need separate evaluation.

\section{Conclusions}\label{sec:conclusions}
This study bridges a critical gap in quantum computational fluid dynamics by developing and benchmarking a tailored, circuit-level HHL implementation for the 1D linear advection equation. Rather than relying on black-box solvers, we constructed an end-to-end pipeline from Qiskit components, enabling direct performance evaluations on IBM's Nighthawk processor. A primary contribution of this work is the adaptation of the HHL protocol to the unique demands of CFD numerical schemes. This includes a normal-equation reformulation to process the non-Hermitian Crank--Nicolson implicit system---which we demonstrate introduces negligible condition-number penalties---and the development of a clock-free Fourier multiplier that bypasses traditional phase estimation, making hardware execution viable on current NISQ devices.

Furthermore, we introduce a robust signed-field and norm reconstruction protocol that effectively translates raw quantum state measurements back into physical macroscopic flow fields utilizing Hadamard-basis interference. While noiseless simulations successfully recovered the transported packet on resolved grids, execution on physical hardware revealed substantial error retention across both explicit and implicit schemes. These findings highlight that algorithmic conditioning alone does not dictate hardware fidelity, and practical speedups remain tightly bounded by current device noise. Ultimately, by providing a validated, end-to-end implementation for implicit time integration, this work establishes a concrete architectural foundation for encoding, executing, and measuring more complex quantum CFD problems.

\section*{Code Availability}
The code and supporting data are publicly available at
\url{https://github.com/batuhanyalcin/quantum-hhl-periodic-advection}.
The repository includes \path{quantum_hhl_corrected.ipynb}, analysis scripts,
specified dependency versions, and reproduction instructions. The project
code and documentation are distributed under the MIT license.

The accompanying \path{supporting_data.zip} contains measurement counts,
job identifiers, submitted native circuits, calibration records, and
the common-time and reconstruction-sensitivity analyses.
The script \path{prepare_data.py} verifies and extracts these materials.
The manuscript figures and tables can be regenerated from the saved data
using \path{analysis/make_overleaf_figures.py} and
\path{analysis/build_overleaf_tables.py}, respectively, without submitting
new quantum hardware jobs.

\begin{samepage}
\section*{Acknowledgments}
We thank IBM Quantum and Rensselaer Polytechnic Institute for access to \texttt{ibm\_rensselaer} (Nighthawk revision~1). Generative AI assistance from Anthropic Claude was utilized for phrasing, grammar, organization, and language revision. The authors assume responsibility for all content, including the scientific claims, code, data, verification of results, and the final manuscript
\end{samepage}

\end{document}